\documentstyle[12pt]{article}
\begin{document}
\begin{flushright}
hep-th/0305118\\
SNBNCBS-2003\\
AS-ICTP-2003
\end{flushright}

\vskip 1.7cm

\begin{center}
{\bf { \large WIGNER'S LITTLE GROUP AND BRST COHOMOLOGY\\
FOR ONE-FORM ABELIAN GAUGE THEORY}}

\vskip 3.5cm

{ R. P. MALIK}
\footnote{ E-mail address: malik@boson.bose.res.in  }\\
{\it S. N. Bose National Centre for Basic Sciences,} \\
{\it Block-JD, Sector-III, Salt Lake, Calcutta-700 098, India} \\

and\\

{\it  The Abdus Salam International Centre for Theoretical Physics, 
Trieste, Italy}\\

\vskip 3cm

\end{center}

\noindent
{\bf Abstract:}
We discuss the (dual-)gauge transformations for the gauge-fixed
Lagrangian density and establish their intimate connection with the translation
subgroup $T(2)$ of the Wigner's little group for the free one-form Abelian
gauge theory in four $(3 + 1)$-dimensions (4D) of spacetime. Though the
relationship between the usual gauge transformation for the Abelian massless
gauge field and $T(2)$ subgroup
of the little group is quite well-known,
such a connection between the dual-gauge transformation and the little
group is a new observation. The above connections
are further elaborated and demonstrated in the framework of
Becchi-Rouet-Stora-Tyutin (BRST) cohomology
defined in the quantum Hilbert space of states where the Hodge decomposition
theorem (HDT) plays a very decisive role. \\


$~~~~~~~~~~$ Wigner's little group\\



\baselineskip=16pt

\newpage

\noindent
{\bf 1 Introduction}\\

\noindent
In the classification scheme of the elementary particles, the
Wigner's little group [1] plays a very important
and decisive role [2]. It was first
Weinberg [2-4] and later Han etal [5-7] who demonstrated a very
interesting connection between the
transformation generated by the Abelian invariant
translation subgroup $T(2)$ of the Wigner's little group and the
$U(1)$ gauge transformation for the one-form ($A = dx^\mu A_\mu$)
Abelian gauge field $A_\mu$ of the Maxwell theory in four
$( 3 + 1)$-dimensions of spacetime. It is a common folklore that the
latter symmetry transformations are generated by the first-class constraints
(in the language of the Dirac's classification scheme [8,9]) of the Abelian
gauge theory which forms the internal $U(1)$  symmetry group of 
transformations. On the contrary, the former transformations are generated 
by the Wigner's little group that constitutes the spacetime symmetry group
of transformations for a given (gauge) theory. In more precise words,
the translation subgroup $T(2)$ of the Wigner's little group
keeps the momentum vector $k_\mu$ of the massless
(i.e. $k^2 = 0$) gauge particle invariant
but changes the polarization vector $e_\mu$ of the (one-form) gauge field in
exactly the same manner as the $U(1)$ gauge transformation generated by the
first-class constraints of the gauge theory. Thus, the Abelian one-form
gauge theory (i.e. free Maxwell theory) provides a fertile ground for the
discussion of the internal symmetry and the spacetime symmetry {\it together}
in a beautiful setting. Recently, in a set of papers [10-13],
the gauge transformations connected with a variety of Abelian gauge theories
have been shown to be connected with the
translation subgroups of the Wigner's little group.

For some of the (non-)interacting gauge theories in two $(1 + 1)$-dimensions
(2D) and 4D of spacetime, it has been found that there exists a discrete
symmetry for the Lagrangian density of the theory which corresponds to
the existence of a specific kind of ``duality'' in the theory. This
duality entails upon the theory to possess (i) a local dual-gauge
symmetry transformation for the Lagrangian density, and (ii) an analogue
of the Hodge duality $*$ operation of differential geometry. Such a class of
(non-)interacting and duality invariant gauge field theories provide a set of
tractable field theoretical models for the Hodge theory where the local,
covariant and continuous symmetry transformations (and the corresponding
generators) are identified with the
de Rham cohomological operators of differential geometry. In this context,
mention can be made of many interesting field theoretical models such
as (i) the free 2D Abelian gauge theory [14-16], (ii) the interacting
2D Abelian gauge theory where there is an interaction between $U(1)$
gauge field and the Dirac fields [17,18], (iii) the self-interacting
2D non-Abelian gauge theory where there is no interaction between the gauge
field and the matter fields [16,19], and (iv) the free Abelian 2-form
gauge theory in 4D [20]. In a recent paper [21], an interesting connection
between the translation subgroup $T(2)$ of the Wigner's little group and
the BRST cohomology has been established
for the free Abelian 2-form gauge theory in 4D. In fact, it is because
of the study of the Wigner's little group that it has been possible to
obtain  the normal mode expansions for the basic fields of the theory [21] that
appear for the consideration of the BRST formalism
(particularly, BRST cohomology) in the framework of Lagrangian formulation.

The purpose of the present paper is to establish a connection between the 
Wigner's little group and the gauge (see, e.g., [22-25]) and the 
dual-gauge symmetry
transformation groups (see, e.g., [26] for details) 
that exist for the free one-form Abelian gauge
theory in four dimensions of spacetime. The salient features of the gauge
and the dual-gauge transformations are (i) the latter transformations are
continuous, non-local and non-covariant whereas the former are continuous,
local and covariant. (ii) It is the gauge-fixing term that remains invariant 
under
the latter transformations. The electric and magnetic fields are left
invariant under the former transformations. (iii) The magnetic field
remains invariant under both the transformations. We demonstrate that both
(the local gauge and the non-local dual-gauge) symmetries owe their
origin to the Wigner's little group as the latter encompasses both the
symmetries in its folds in a subtle way. Furthermore, we show that the
(dual-)gauge (or (co-)BRST)
transformed physical states are found to be the sum of the
original physical states and the BRST (co-)exact states. Thus, the increment
in the physical state due to the (dual-)gauge
(or (co-)BRST) transformations turns out
to be a cohomologically trivial state. For this proof, we exploit (i)
the HDT in the quantum Hilbert space of states
(QHSS), and (ii) choose the physical state to be the harmonic state of the
Hodge decomposed state in the QHSS. The choice of the harmonic state to be
the physical state is guided by some aesthetic reasons because this state is
the most symmetrical nontrivial state which is (anti-)BRST invariant as well as
(anti-)co-BRST invariant, simultaneously. One of the most crucial points
of our whole discussion is the choice of the momentum vectors
$k_\mu = (\omega, 0, 0, - \omega)^T$ and $k^\mu = (\omega, 0, 0, \omega)^T$
for the massless ($k^2 = 0$) photon, propagating along the z-direction of the
4D spacetime manifold with energy $\omega$. This choice enables us to get a 
simple
expression for the non-local and  non-covariant dual-gauge (or co-BRST)
transformations in the phase space. In fact, the ugly features
of non-locality and non-covariance disappear for this choice.
Furthermore, this choice of the reference frame allows us to get the {\it same}
physical inferences from the conserved and nilpotent
BRST and co-BRST charges when they apply on the physical harmonic
state in the requirement of physicality criteria. This unique feature is
not present in our earlier works [14-19] where the BRST and co-BRST charges
lead to {\it different} physical consequences when they are applied on
a single photon state for the 2D gauge theories. To be more precise, the
BRST charge implies the transversality condition on the 2D photon but the
co-BRST charge leads to the ``dual'' transversality condition between
the momentum vector $k_\mu$ and the polarization vector $e_\mu$. An exact
generalization of these results has been obtained for the free 2-form
Abelian gauge theory in 4D with conserved and nilpotent 
(co-)BRST charges [20,21].

Our present study is essential primarily on three counts. First and foremost,
to the best of our knowledge,  the derivation of the connection between the 
continuous, {\it non-local} and
{\it non-covariant} dual-gauge (or co-BRST) transformations and the Wigner's
little group is a new result which is 
different from the well-known connection between the usual continuous, 
{\it local} and {\it covariant} $U(1)$ gauge
transformations for the massless 1-form gauge field and the $T(2)$ subgroup of
the little group. Second, the free Maxwell $U(1)$ Abelian gauge theory is 
one of the simplest 1-form gauge theories where the gauge and dual-gauge 
symmetry transformations co-exist together for the gauge-fixed Lagrangian 
density of the theory. Thus, the study of 
the deeper reasons for their existence in the framework of the
translation subgroup $T(2)$ of the Wigner's little group is the first
step towards our main goal of understanding the more complicated (e.g.
2-form, 3-form, etc.) gauge theories which have relevance in the context of
(super)string theories and their close cousins D-branes.
Finally, to corroborate the above {\it assertions},
it is worthwhile to state that, in the context of the 4D free Abelian
2-form gauge theory, it has been shown (see, e.g., [20,21] for details)
that, in some sense, the transformations generated by the little group
on the polarization tensor are more fundamental than the corresponding 
changes brought about by the first-class constraints of the
gauge and dual-gauge symmetry groups. This is because
of the fact that, only when we demand the {\it consistency} of the 
transformations generated by the usual (dual-)gauge
groups with that of the little group, do we get a set of certain specific 
relationships between the 
parameters of the little group and the (dual-)gauge groups. This relationship, 
finally, entails upon the (anti-)ghost fields of the BRST formalism
to obey certain specific restrictions. This, in turn, allows us to obtain
the normal mode expansions for the ghost fields (see, [21] for details)
which play very important roles in the proof of the quasi-topological nature
of the free 2-form Abelian gauge theory in the framework of BRST 
cohomology [20,21]. Thus, the study of the Wigner's little group does shed 
some light on the formal aspects of the gauge field theories and their
discussion in the framework of BRST formalism. It is well-known that
the 2-form Abelian gauge fields are important in the context of
(super)string theories, D-branes and noncommutative geometry.

The material of our present paper is organized as follows. In section 2,
we briefly discuss the (dual-)gauge symmetry transformations for the
gauge-fixed Lagrangian density of the free Abelian (one-form) gauge theory
and show that the restrictions on the
(dual-)gauge parameters are similar. These transformations are upgraded
to the nilpotent (co-)BRST transformations in section 3.
The central of the present paper are sections 4 and 5
where we show the connection
between the Wigner's little group and the (dual-)gauge transformations and
comment on such relationship in the language of the BRST cohomology
where the HDT in the QHSS plays a very decisive role.
Finally, we make some concluding remarks in section 6 and point out a few
future directions that can be pursued later.\\

\noindent
{\bf 2 (Dual-)gauge transformations}\\

\noindent
Let us start off with the gauge-fixed
Lagrangian density ${\cal L}_{0}$ for the four
$(3 + 1)$-dimensional free Abelian gauge theory in the Feynman gauge
(see, e.g., [22-25])
$$
\begin{array}{lcl}
{\cal L}_{0} = - \frac{1}{4} F^{\mu\nu} F_{\mu\nu} - \frac{1}{2}
(\partial \cdot A)^2 \equiv \frac{1}{2} ({\bf E^2 - B^2}) - \frac{1}{2}
(\partial \cdot A)^2,
\end{array}\eqno(2.1)
$$
where $F_{\mu\nu} = \partial_\mu A_\nu - \partial_\nu A_\mu$ is the
anti-symmetric second-rank curvature tensor (with $ F_{0i} = E_i
\equiv {\bf E}$ = electric field, $B_i = \frac{1}{2}\;\epsilon_{ijk} F_{jk}
\equiv {\bf B}$ = Magnetic field)
defined through the 2-form $F = d A = \frac{1}{2} (dx^\mu
\wedge dx^\nu) (F_{\mu\nu})$.  As is evident, this 2-form is derived by the
application of the exterior derivative $ d = dx^\mu \partial_\mu$
(with $ d^2 = 0$) on the connection one-form $A = dx^\mu A_\mu$
which defines $A_\mu$ as the vector potential 
\footnote{ We adopt here the conventions and notations in such a way that the
flat 4D Minkowski spacetime metric
$\eta_{\mu\nu} =$ diag $(+1, -1, -1, -1)$ and the totally antisymmetric
Levi-Civita tensor obeys $\varepsilon_{\mu\nu\kappa\sigma}
\varepsilon^{\mu\nu\kappa\sigma} = - 4!, \; \varepsilon_{\mu\nu\kappa\sigma}
\varepsilon^{\mu\nu\kappa\eta} = - 3! \delta^\eta_\sigma, \mbox {etc.}, \;
\varepsilon_{0123} = + 1 = - \varepsilon^{0123}, \; \varepsilon_{0ijk} =
\epsilon_{ijk}$ and $\Box = (\partial_{0})^2 - (\partial_{i})^2
\equiv (\partial_{0})^2 - (\nabla)^2$. Here the Greek indices 
$\mu, \nu, \kappa.....= 0, 1, 2, 3$
correspond to the spacetime directions on the 4D manifold and the Latin
indices $ i, j, k.....= 1, 2, 3$ stand only for the space directions on
the 3D submanifold. The
3-vectors on the submanifold are represented by the bold faced letters.}.
The gauge-fixing term
$(\partial \cdot A) = (- * d * A)$ is defined through the application of
the dual-exterior derivative $\delta = - * d *$ (with $ \delta^2 = 0$) on
the one-form $A$. Here the $*$ operation is the
Hodge duality operation on the 4D spacetime manifold. The application of the
Laplacian operator
$\Delta = (d + \delta)^2 = d \delta + \delta d$ on the one-form $A$ leads to
$\Delta A = dx^\mu \Box A_\mu$. In fact, the equation of motion ($\Box A_\mu
= 0$), emerging from the above gauge-fixed Lagrangian density, is captured by
the Laplacian operator in the sense that
it (i.e. $\Box A_\mu = 0$) can be derived by
the  requirement of the validity of the Laplace equation $\Delta A = 0$.
Together the set of geometrical operators
$(d,\delta,\Delta)$ define the de Rham cohomological
properties of the differential forms and obey the algebra:
$ d^2 = \delta^2 = 0, \Delta = (d + \delta)^2 = \{ d , \delta \},
[ \Delta, d ] = [ \Delta, \delta ] = 0$ (see, e.g., [27-30]).  It is
unequivocally clear that both the terms of the above Lagrangian density have
deep connections with the key cohomological operators of the differential
geometry. Their invariances, therefore, will play some prominent roles in our
whole discussions about the (dual-)gauge 
transformations as well as the corresponding (dual-)BRST transformations. 
In fact, the nomenclature of the gauge (or BRST) and 
the dual-gauge (or dual(co)-BRST) symmetry transformations owes
its origin to the {\it invariances} of the above terms.

It is straightforward to check that the above Lagrangian density, under
the following local $U(1)$ gauge [22-25] and dual-gauge transformations 
(see, e.g., [26])
$$
\begin{array}{lcl}
A_\mu (x) &\rightarrow&  A_\mu^{(g)} (x) = A_\mu (x)
+ \partial_\mu\; \alpha (x), \nonumber\\
A_0 (x) &\rightarrow& A_0^{(dg)} (x) = A_0 (x) + i \;\beta (x), \nonumber\\
A_i (x) &\rightarrow& A_{i}^{(dg)} (x) = A_i (x) +
i\; {\displaystyle \frac{\partial_{0} \partial_{i}}{\nabla^2}} \;\beta (x),
\end{array}\eqno (2.2)
$$
remains invariant if the parameters of the transformations are restricted
to obey $\Box \alpha = 0, \Box \beta = 0$.
Under the infinitesimal version of the above (dual-)gauge
transformations $\delta_{(d)g}$, the following changes occur
$$
\begin{array}{lcl}
&&\delta_{g} A_\mu = \partial_\mu \alpha, \qquad \delta_{g}  E_i = 0,
\qquad \delta_{g} B_i = 0, \qquad \delta_{g} (\partial \cdot A)
= \Box \alpha, \nonumber\\
&&\delta_{dg} A_0 = i \beta, \qquad
\delta_{dg} A_i = i {\displaystyle \frac{\partial_0 \partial_i}
{\nabla^2} \;\beta },
\qquad \delta_{dg} B_i  = 0, \nonumber\\
&& \delta_{dg} (\partial \cdot A)= 0, \qquad
\delta_{dg}  E_i = i \Bigl ( {\displaystyle \frac{\partial_0 \partial_0}
{\nabla^2}} - 1 \Bigr )\; \partial_i \beta
\equiv i {\displaystyle \frac{\Box} {\nabla^2}} \partial_i \beta.
\end{array} \eqno(2.3)
$$
Some of the relevant points, at this stage, are as follows.
First, it is the kinetic energy term (more precisely the 2-form
curvature tensor $F_{\mu\nu}$ itself) and the gauge-fixing term
(more precisely $(\partial \cdot A)$ itself) that remain invariant
under the gauge and dual-gauge transformations, respectively. Second,
exactly the same
restrictions (i.e. $\Box \alpha = \Box \beta = 0$) are imposed on the
(dual-)gauge parameters for the invariance of the Lagrangian density under
the (dual-)gauge transformations.
Finally, the latter transformations in (2.2) are christened as the
dual-gauge transformations because $(\partial \cdot A)$ and
$F_{\mu\nu}$ are `Hodge-dual' to each-other from the point of view of
their derivation using the operation of the
(co-)exterior derivatives ($\delta$ and $d$) on the one-form 
$A = dx^\mu A_\mu$ defining the vector (gauge) potential.

Using the restriction $ \Box \beta = 0 \rightarrow \partial_{0} \partial_{0}
\beta = \nabla^2 \beta$ as an input, it can be checked that the above
dual-gauge
transformations on the vector field $A_\mu$ can be re-expressed as [31]
$$
\begin{array}{lcl}
\tilde \delta_{dg} A_\mu (x) = i \partial_\mu \Bigl ( {\displaystyle
\frac{\partial_{0}} {\nabla^2}} \beta (x) \Bigr ),
\end{array} \eqno(2.4)
$$
which imply $ \delta_{dg} A_{0} = i \beta, \delta_{dg} A_i = i (\partial_{0}
\partial_{i}/ \nabla^2) \beta$ as is the case in (2.3) only when
$\partial_{0} \partial_{0} \beta = \nabla^2 \beta$ is used explicitly.
It will be noted,
however, that the above form of the dual-gauge transformation
$\tilde \delta_{dg}$ does not keep the gauge-fixing term invariant
(i.e. $\tilde \delta_{dg} (\partial \cdot A) \neq 0$). Thus,
we shall {\it not} use both the forms of non-local dual-gauge
transformations (cf. (2.3) and (2.4)) for our later discussions in
sections 4 and 5. We shall focus on the transformations (2.3) only
for its generalization to the co-BRST symmetry transformations (as the 
gauge-fixing term remains invariant under it). There are more general 
discussions [31] on the
non-covariance and non-locality of the transformations (2.3) in the framework
of BRST formalism. However, the generalized BRST-type symmetries turn out
to be nilpotent only for a specific value of the parameter
of the theory [31]. Thus, we shall avoid deeper discussions on the symmetry
transformations of [31] and concentrate only on the continuous,
(non-)local and (non-)covariant symmetry transformations discussed in [26].\\

\noindent
{\bf 3 (Co-)BRST symmetries}\\

\noindent
The gauge-fixed Lagrangian density 
${\cal L}_{0}$ of equation (2.1) can be generalized to the BRST
invariant Lagrangian density ${\cal L}_{b}$ as (see, e.g., [22-25])
$$
\begin{array}{lcl}
{\cal L}_{b} = - \frac{1}{4} F^{\mu\nu} F_{\mu\nu} - \frac{1}{2}
(\partial \cdot A)^2 - i \partial_\mu\; \bar C \; \partial^\mu \; C,
\end{array}\eqno(3.1)
$$
where the anticommuting ($\bar C^2 = C^2 = 0, C \bar C + \bar C C = 0$)
(anti-)ghost fields $(\bar C) C$ are required in the theory to maintain
the unitarity and ``quantum'' gauge (i.e. BRST) invariance together
at any arbitrary order of perturbative calculations (see, e.g., [32]). The
above Lagrangian density (3.1) respects the following on-shell
($ \Box C = \Box \bar C = 0$) nilpotent ($ s_{(a)b}^2 = 0$) (anti-)BRST
$s_{(a)b}$ (with $ s_b s_{ab} + s_{ab} s_{b} = 0$) symmetry transformations
\footnote{ We follow here the notations and conventions of Weinberg [22].
Actually, in its full glory, the nilpotent $(\delta_{(A)B}^2 = 0)$
(anti-)BRST transformations ($\delta_{(A)B}$)
are the product ($\delta_{(A)B} = \eta s_{(a)b}$)
of an anticommuting ($ \eta C + C \eta = 0, \eta \bar C + \bar C \eta = 0$)
spacetime independent parameter $\eta$ and $s_{(a)b}$ with $s_{(a)b}^2 = 0$.}
$$
\begin{array}{lcl}
&&s_{b} A_\mu = \partial_\mu C, \qquad s_{b} C = 0, \qquad s_{b} \bar C
= - i (\partial \cdot A), \nonumber\\
&&s_{ab} A_\mu = \partial_\mu \bar C, \qquad s_{ab} \bar C = 0, \qquad
s_{ab} C = + i (\partial \cdot A).
\end{array} \eqno(3.2)
$$
The salient features, at this juncture, are (i) the physical fields
$E_i$ and $B_i$ remain invariant ($s_{(a)b} E_i = 0, s_{(a)b} B_i = 0$)
under the (anti-)BRST transformations. (ii) The (anti-)BRST transformations
are the generalization of the gauge transformations of (2.3) without
the restriction like $\Box \alpha = 0$. (iii) The bosonic
gauge parameter $\alpha$
has been replaced by $\alpha = \eta \bar C$ and $ \alpha = \eta C$ for the
derivation of the (anti-)BRST transformations.

The non-local and non-covariant dual gauge transformations of (2.3) can
be elevated to the on-shell ($\Box C = \Box \bar C = 0$) nilpotent
($s_{(a)d}^2 = 0)$ symmetry transformations by replacing the dual
gauge parameter $ \beta = \eta C$ and $\beta = \eta \bar C$ corresponding
to the (anti-)dual BRST (i.e. (anti-)co-BRST) $s_{(a)d}$ symmetry 
transformations
(with $s_{d} s _{ad} + s_{ad} s_{d} = 0$) as
$$
\begin{array}{lcl}
&&s_{d} A_0 = i \bar C, \qquad s_{d} A_i = i
{\displaystyle \frac{\partial_0 \partial_i} {\nabla^2}} \bar C,
\qquad s_{d} \bar C = 0, \qquad s_{d} C = - A_0 +
{\displaystyle
\frac{\partial_0 \partial_i} {\nabla^2}} A_i, \nonumber\\
&&s_{ad} A_0 = i C, \qquad s_{ad} A_i = i
{\displaystyle \frac{\partial_0 \partial_i} {\nabla^2}} C,
\qquad s_{ad} C = 0, \qquad s_{ad} \bar C = A_0 -
{\displaystyle
\frac{\partial_0 \partial_i} {\nabla^2}} A_i.
\end{array} \eqno(3.3)
$$
The key and relevant points, at this stage, are (i) it is the gauge-fixing
term that remains invariant ($s_{(a)d} (\partial \cdot A) = 0$)
under the (anti-)co-BRST transformations $s_{(a)d}$. (ii) A single dual
gauge symmetry
transformations $\delta_{dg}$ in (2.3) leads to the existence of a
couple of nilpotent (anti-)co-BRST symmetry transformations. (iii) The
physical magnetic field $B_i$ remains invariant
(i.e. $s_{(a)d} B_i = 0$) under $s_{(a)d}$ but
the electric field $E_i$ transforms ($s_d E_i = i (\Box/\nabla^2)
\partial_i \bar C, s_{ad} E_i = i (\Box/ \nabla^2) \partial_i C$)
under $s_{(a)d}$ in such a way as to cancel the contributions coming from
the transformation of the ghost term under $s_{(a)d}$. It is obvious that
a bosonic symmetry $s_{w}$ (with $s_{w}^2  \neq 0$) can be obtained from the
anticommutator of the nilpotent symmetries  (i.e. $s_{w} = \{ s_{b}, s_{d} \}
= \{ s_{ad}, s_{ab} \}$). However, for our discussions, this symmetry is
not required in its full glory. Some elementary discussions about it can be
found in [26].

The generators for the above (non-)local, continuous, (non-)covariant and 
nilpotent (co-)BRST transformations can be computed from the Noether 
conserved current as listed below
$$
\begin{array}{lcl}
Q_{d} &=& {\displaystyle \int} d^3 x\;
\bigl [ \;
{\displaystyle \frac{ \partial_{0} (\partial \cdot A)} {\nabla^2}}
\dot {\bar C} - (\partial \cdot A) \bar C
\;\bigr ], \nonumber\\
Q_{b} &=& {\displaystyle \int} d^3 x\;
\bigl [\; \partial_{0} (\partial \cdot A) C - (\partial \cdot A) \dot C
\;\bigr ].
\end{array} \eqno(3.4)
$$
From the above expressions, the conserved and nilpotent charges corresponding
to the anti-BRST and anti-co-BRST transformations can be computed by the
replacement $ C \rightarrow \pm i \bar C, \bar C \rightarrow \pm i C$ which
turns out to be the discrete symmetry transformation for the ghost part
of the Lagrangian density (3.1). For any generic field $\Phi
= A_\mu, C, \bar C$ of the theory, the (anti-)BRST and (anti-)co-BRST
transformations of (3.2) and (3.3) can be succinctly expressed as
$$
\begin{array}{lcl}
s_{r} \Phi = - i \; [ \Phi, Q_{r} ]_{\pm}, \;\qquad\; r = b, ab, d, ad, w, g,
\end{array} \eqno(3.5)
$$
where the subscripts $(+)-$ on the brackets correspond to the (anti-)commutators
for the generic field $\Phi$ being (fermionic)bosonic in nature and $Q_{w}$
(i.e. $ Q_{w} = \{ Q_{b}, Q_{d} \} = \{ Q_{ad}, Q_{ab} \}$)
is the generator for the bosonic symmetry transformation and $Q_{g}$ is the
ghost conserved charge that corresponds to the infinitesimal scale symmetry
transformations $s_{g} C = - \lambda C, s_{g} \bar C = + \lambda \bar C,
s_{g} A_\mu = 0$ under which the action remains invariant. Here the
transformation parameter $\lambda$ is global. Thus, in total, there are
six continuous, (non-)local and (non-)covariant  symmetries in the theory.\\

\noindent
{\bf 4 Wigner's little group and gauge transformations}\\

\noindent
The Wigner's little group corresponds to the maximal subgroups of the
Lorentz group that leave
the four momenta of the free relativistic particles invariant. The internal
symmetry properties associated with the gauge particle are captured by
the little group which turns out to be locally isomorphic to the three 
dimensional
rotation group and the two-dimensional Euclidean group (see, e.g., [33]
for details).
The most general form of the Wigner's little
group matrix $\{ W^\mu_\nu (\theta, u, v) \}$ for a massless
(gauge) particle moving along the $z$-direction of the 4D spacetime manifold
is [2-7]
$$
\begin{array}{lcl}
\{ W (\theta, u, v) \} =
\left ( \begin{array}{cccc}
\bigl (1 + {\displaystyle \frac{u^2 + v^2} {2}} \bigr ) &
\bigl (u cos \theta -  v sin \theta \bigr )
& \bigl (u sin \theta + v cos \theta \bigr ) & -
\bigl ({\displaystyle \frac{u^2 + v^2}{2}} \bigr )\\
u & cos \theta & sin \theta  & - u \\
v & - sin \theta & cos \theta & - v\\
\bigl ({\displaystyle \frac{u^2 + v^2}{2}} \bigr ) &
\bigl (u cos \theta - v sin \theta \bigr )
&\bigl  (u sin \theta + v cos \theta \bigr ) &
\bigl (1 - {\displaystyle \frac{u^2 + v^2} {2}} \bigr ) \\
\end{array} \right ),
\end{array} \eqno(4.1)
$$
where $\theta$ is the rotational parameter and $u, v$ are the translational
parameters defining $T(2)$ in the $xy$ plane. By definition, this matrix
preserves the four momentum $k^\mu = (\omega, 0, 0, \omega)^T$ of
a massless ($k^2 = 0$) (gauge) particle with energy $\omega$ and it can be
factorized elegantly into 3D rotation group and 2D Euclidean group. Both
these statements can be expressed mathematically, in a combined fashion,  as
$$
\begin{array}{lcl}
(k^\mu) \rightarrow (k^\mu)^\prime
= W^\mu_\nu\; (k^\nu) = (k^\mu), \; \qquad
W (\theta, u. v) = R (\theta)\; W (0, u, v).
\end{array} \eqno(4.2)
$$
The matrix $R(\theta)$ in the above represents the rotation about the z-axis
$$
\begin{array}{lcl}
R (\theta) =
\left ( \begin{array}{cccc}
1 & 0 & 0 & 0 \\
0 & cos \theta & sin \theta & 0 \\
0 & - sin \theta  & cos \theta & 0 \\
0 & 0 & 0 & 1 \\
\end{array} \right ),
\end{array} \eqno(4.3)
$$
and the matrix $ \{ W(0,u,v) \}$ is found to be isomorphic
to the two parameter translation group $T(2)$ (i.e. $T(2) \sim W (0,u,v)$)
in the two-dimensional Euclidean plane
($xy)$ which is a plane perpendicular to the propagation of the light-like
(massless) particle along the z-direction of the 4D spacetime manifold.

It is obvious from the Lagrangian density (3.1) that the equations of motion
for the basic fields are: $ \Box A_\mu = \Box C = \Box \bar C = 0$. These
imply the masslessness ($k^2 = 0$) of the photon as well as the (anti-)ghost
fields. Thus, the choice $k^\mu = (\omega, 0 , 0, \omega)^T$ is consistent
with the equations of motion which imply $k^2 = 0$. The transversality
\footnote{In the next section, it will be seen that this condition
(i.e. $ k \cdot e = 0$) emerges
automatically due to the requirement that the physical states (i.e. $|phys>$)
of the theory are annihilated (i.e. $Q_{b}\; |phys> = 0$)
by the conserved ($ \dot Q_{b} = 0$) and nilpotent
($ Q_{b}^2 = 0$) BRST charge $Q_{b}$ due to the physicality criteria.}
$ k_\mu e^\mu = 0$ of photon
implies that the polarization vector $e^\mu (k)$ of the photon can be
chosen to be $e^\mu (k) = (0, e^1 (k), e^2 (k), 0)^T$. The $U(1)$ gauge
transformation of (2.2) (generated by the first-class constraints
of the theory) can be exploited to express itself
in terms of the transformation on $e^\mu (k)$ as (see, e.g., [10])
$$
\begin{array}{lcl}
e^\mu (k) \rightarrow (e^\mu)^{(g)} (k) = e^\mu (k) + i k^\mu \alpha (k).
\end{array} \eqno(4.4)
$$
It will be noted that, in general, one can choose $e^\mu (k)
= (e^0, e^1, e^2, e^0)^T (k)$ which will be consistent with the transversality
condition ($k_\mu e^\mu = 0$) with our choice
of the reference frame where $k_\mu = (\omega, 0, 0, - \omega)^T$. 
However, it can be seen, using the gauge
transformation (2.2), that the component $e^0 (k)$ of the polarization
vector $e^\mu (k)$ can be gauged away
(i.e. $(e^0)^{(g)} = 0$) by the choice $\alpha (k) = - (e^{(0)}/ i \omega)$
in the gauge transformation
$$
\begin{array}{lcl}
e^0 (k) \rightarrow (e^0)^{(g)} (k) = e^0 (k) + i \omega \alpha (k).
\end{array} \eqno(4.5)
$$
Thus, ultimately, we end up with $e^\mu (k) = (0,  e^1,  e^2, 0)^T (k)$. Now,
concentrating on the role of the translation subgroup $T(2) \sim W (0, u, v)$
in generating the $U(1)$ gauge transformation on $e^\mu (k)$, it can be
readily checked that
$$
\begin{array}{lcl}
e^\mu (k) \rightarrow (e^\mu)^{\prime} (k) = W^\mu_\nu (0, u, v)\; e^\nu
\equiv e^\mu (k) +
\Bigl ({\displaystyle \frac{u e^1 + v e^2} {\omega}} \Bigr )\; k^\mu.
\end{array} \eqno(4.6)
$$
It is unequivocally clear that both the transformations in (4.4) and (4.6)
are identical for the following relationship between the infinitesimal
gauge parameter $\alpha (k)$ and the parameters $u$ and $v$
of the translation subgroup $T(2)$ of the Wigner's little group
$$
\begin{array}{lcl}
\alpha (k) =
{\displaystyle \frac{u e^1 + v e^2} {i \omega}}.
\end{array} \eqno(4.7)
$$

Now, let us focus on the dual-gauge transformations of (2.2). Following
the prescription of [10], these transformations in the momentum phase space
can be expressed, in terms of the polarization vector components, as
$$
\begin{array}{lcl}
e^0 (k) &\rightarrow& (e^0)^{(dg)} (k) = e^0 (k) + \;i \;\beta (k), \nonumber\\
e^i (k) &\rightarrow& (e^i)^{(dg)} (k) = e^i (k) + i \;
{\displaystyle \frac{k^0 k^i} {{\bf k}^2}} \; \beta (k).
\end{array} \eqno(4.8)
$$
Concentrating on our choice of the reference frame where the momentum vector
$k^\mu = (\omega, 0, 0, \omega)^T$, the above transformations can be
re-expressed as
$$
\begin{array}{lcl}
e^0 (k) &\rightarrow& (e^0)^{(dg)} (k) = e^0 (k) + i \beta (k), \nonumber\\
e^1 (k) &\rightarrow& (e^1)^{(dg)} (k) = e^1 (k), \qquad
e^2 (k) \rightarrow (e^2)^{(dg)} (k) = e^2 (k), \nonumber\\
e^3 (k) &\rightarrow& (e^3)^{(dg)} (k) = e^3 (k) + i \beta (k).
\end{array} \eqno(4.9)
$$
It is obvious, at this stage, that (i) the transverse components of the 
polarization vector do not transform at all under the dual-gauge 
transformation for the choice
$k^\mu = (\omega, 0, 0, \omega)^T$ and $k_\mu = (\omega, 0, 0, -\omega)^T$,
and (ii) the scalar and longitudinal components transform in exactly the 
same manner. This observation should be contrasted with the gauge 
transformations where the scalar
and longitudinal components (with $ e^0 = e^3$) can be gauged away and the
transverse components transform.
Finally, the above transformation (4.9) can be concisely written in the four
vector notation as
$$
\begin{array}{lcl}
e^\mu (k) \rightarrow (e^\mu)^{(dg)} (k) = e^\mu (k) +
{\displaystyle \Bigl (\frac{i \beta (k)} {\omega}} \Bigr )\; k^\mu.
\end{array} \eqno(4.10)
$$
It is obvious that both the transformations in (4.10) and (4.6)
are identical for the following relationship between the
infinitesimal dual-gauge parameter
$\beta (k)$ and the parameters $u$ and $v$
of the translation subgroup $T(2)$ of the Wigner's little group
$$
\begin{array}{lcl}
\beta (k) =
{\displaystyle \frac{u e^1 + v e^2} {i}}.
\end{array} \eqno(4.11)
$$
This establishes the fact that (dual-)gauge transformations owe their origin
to the transformations generated by the translation subgroup $T(2)$ of
the Wigner's little group. In this analysis and treatment,
the parameters of the (dual-)gauge transformations are chosen in 
terms of the parameters of the translation subgroup $T(2)$ as given in (4.11)
and (4.7). It is interesting to note that, for the above choice of the reference
frame in the context of the discussion on little group, 
the infinitesimal (dual-)gauge
transformation parameters of $A_\mu$ differ by a factor of the energy $\omega$
(i.e. $ \beta = \omega \alpha$) of the photon (cf. (4.7) and (4.11)). 
It is
worthwhile to emphasize at this juncture that, for the free 2-form Abelian
gauge theory, the distinction between the gauge and dual-gauge transformations
is quite clear and lucid [20,21]. Furthermore, the trivial relationship
like $\beta = \omega \alpha$ does not exist in the case of the 2-form
Abelian gauge theory (see, e.g., [21] for details).\\

\noindent
{\bf 5 Wigner's little group and BRST cohomology}\\

\noindent
Here we shall recapitulate some of the key
and pertinent points of the discussion connected with the BRST cohomology
by Weinberg [22] for the gauge transformations. To
this end in mind, we first express the normal mode expansion for the
basic fields ($A_\mu, C, \bar C)$ of the Lagrangian density (3.1) in the
(momentum) phase space as
$$
\begin{array}{lcl}
A_\mu (x) &=& {\displaystyle \int \; \frac{d^3 k} {(2\pi)^{3/2} (2 k^0)^{3/2}}}
\;\bigl [\; a_{\mu}^\dagger (k) \;e^{i k \cdot x} \;+ \;
a_{\mu} (k) \; e^{- i k \cdot x}
\; \bigr ], \nonumber\\
C (x) &=& {\displaystyle \int \; \frac{d^3 k} {(2\pi)^{3/2} (2 k^0)^{3/2}}}
\;\bigl [ \;c^\dagger (k) \;e^{i k \cdot x}\; + \;c (k) \; e^{- i k \cdot x}
\; \bigr ], \nonumber\\
\bar C (x) &=& {\displaystyle \int \; \frac{d^3 k} {(2\pi)^{3/2}
(2 k^0)^{3/2}}} \; \bigl [\; \bar c^\dagger (k) \;e^{i k \cdot x}
\;+\; \bar c (k) \; e^{- i k \cdot x}
\; \bigr ].
\end{array} \eqno(5.1)
$$
The above expansions correspond to the equations of motion 
$\Box A_\mu = \Box C = \Box
\bar C = 0$ obeyed by the basic fields of the theory. Here $k_\mu$ are the
4D momenta and $ d^3 k = d k_{1} d k_2 d k_3$ is the volume in the
momentum space. All the dagger operators are the creation operators and
the non-dagger operators correspond to the
annihilation operators for the basic quanta of the fields. The on-shell
nilpotent version of the BRST symmetries (3.2) can be expressed,
due to (3.5), in terms of the (anti-)commutators with $Q_{b}$
as (see, e.g., [22,16] for details)
$$
\begin{array}{lcl}
&& [ Q_{b}, a_{\mu}^\dagger (k) ] = k_\mu \;c^\dagger (k), \qquad
[ Q_{b}, a_{\mu} (k) ] = - k_\mu c (k), \nonumber\\
&& \{ Q_{b}, c^\dagger (k) \} = 0, \;\;\;\;\;\qquad \;\;\;\;\;
\{ Q_{b}, c (k) \} = 0, \nonumber\\
&& \{ Q_{b}, \bar c^\dagger (k) \} =  i \;k^\mu \;a_{\mu}^\dagger (k), \qquad
\{ Q_{b}, \bar c (k) \} =  - i \;k^\mu \;
a_\mu (k).
\end{array} \eqno(5.2)
$$
Similar kinds of (anti-)commutation relations can be obtained with the
anti-BRST generators but we do not require them for our present
discussions. For aesthetic reasons, we can define the most symmetric
physical vacuum ($|vac>$) of the present theory as
$$
\begin{array}{lcl}
&&Q_{(a)b} \;| vac > = 0, \qquad Q_{(a)d}\; | vac > = 0,
\qquad Q_{w} \;|vac > = 0,
\nonumber\\
&& a_{\mu} (k)\; | vac > = 0, \qquad c (k)\; | vac > = 0, \qquad \;\;
\bar c (k)\; |vac > = 0,
\end{array}\eqno(5.3)
$$
where $Q_{w} = \{ Q_{(a)b}, Q_{(a)d} \}$ is the generator for the
bosonic symmetry transformation that we commented on earlier.
In the above, it is clear that the physical vacuum is (anti-)BRST and
(anti-)co-BRST invariant which imply the invariance w.r.t. $Q_{w}$ as well.
It is lucid and clear now that
a single photon state with polarization $e_\mu (k) $ and momenta $k_\mu$ can be
created from the physical vacuum by the application of a creation operator
$a_\mu^\dagger (k)$ as [22,16]
$$
\begin{array}{lcl}
e^\mu a_{\mu}^\dagger (k)\; | vac > \equiv | e, vac >, \qquad
k^\mu a_{\mu}^\dagger (k)\; | vac > \equiv | k, vac > = -  i \{ Q_{b},
\bar c^\dagger (k) \} \;|vac>,
\end{array} \eqno(5.4)
$$
where the latter state $ |k, vac>$ with momenta $k_\mu$ has been expressed by
exploiting the anti-commutator $\{ Q_{b}, \bar c^\dagger (k) \} = i k^\mu
a_\mu^\dagger (k)$ from (5.2). Exploiting the $U(1)$ gauge transformation
(4.4) on the polarization vector $e_\mu (k) \rightarrow (e_\mu)^{(g)} (k) =
e_\mu (k) + i A k_\mu$, where $A$ is a complex number (which can be
expressed taking the help of (4.4) and (4.7) as
$ A = (u e^1 + v e^2)/ (i  \omega)$), it is straightforward
to check that
$$
\begin{array}{lcl}
| e + i\; A\; k, vac > = | e, vac > + \;
Q_{b}\; (A\; \bar c^\dagger (k)) | vac >, \qquad Q_{b}\; |vac> = 0.
\end{array} \eqno(5.5)
$$
It should be noted that the arbitrariness of $A$ in the gauge
transformation, for the discussion of the BRST cohomology, is being traded
with the arbitrariness of the parameters of the 
Wigner's little group and the energy $\omega$ of the massless field.
This key observation is due to the consistency between 
the transformations generated by the Wigner's little group
and the gauge group.
We conclude from (5.5) that a gauge transformed state corresponding to
an original single photon state (i.e. $ e^\mu (k) a_\mu^\dagger (k) |vac>
\equiv |e, vac>$
with the polarization vector $e_\mu (k)$) is equal to the sum of the original
state $| e, vac>$ plus a BRST exact state. In more sophisticated language,
the gauge transformed state and the original state belong to the same
cohomology class w.r.t. the conserved and nilpotent BRST charge $Q_{b}$. 
In other words, the increment
in the original physical state due to the gauge (or BRST) transformation
is a cohomologically trivial state. Thus, the truly physical state remains
invariant under the gauge (or BRST) transformations (modulo a cohomologically
trivial state which is a BRST exact state).

Now let us focus on the dual-BRST transformations (3.3). These transformations
can be expressed in terms of the conserved and nilpotent dual BRST charge
$Q_{d}$ and creation and annihilation operators of (3.1), by exploiting
the general expression for the transformation (3.5), as
$$
\begin{array}{lcl}
&& [ Q_{d}, a_{0}^\dagger (k) ] =
\bar c^\dagger (k), \qquad
[ Q_{d}, a_{0} (k) ] = \bar c (k), \nonumber\\
&& [ Q_{d}, a_{i}^\dagger (k) ] =
{\displaystyle \frac{k_{0} k_{i}}{{\bf k}^2}}\;
\bar c^\dagger (k), \qquad
[ Q_{d}, a_{i} (k) ] =
{\displaystyle \frac{k_{0} k_{i}}{{\bf k}^2}}\;
\bar c (k), \nonumber\\
&& \{ Q_{d}, c^\dagger (k) \} = i
\Bigl ( a_{0}^\dagger (k) - {\displaystyle \frac{ k_{0} k_{i}} {{\bf k}^2}}
a_{i}^\dagger (k) \Bigr ), \qquad
\{ Q_{d}, \bar c^\dagger (k) \} =  0,
\nonumber\\
&& \{ Q_{d}, c (k) \} =  i
\Bigl ( a_{0} (k) - {\displaystyle \frac{ k_{0} k_{i}} {{\bf k}^2}}
a_{i} (k) \Bigr ), \qquad
\{ Q_{d}, \bar c (k) \} =  0.
\end{array} \eqno(5.6)
$$
Exploiting our choice of the contravariant and covariant
momentum vectors $k^\mu = ( \omega, 0, 0, \omega)^T$ and
$k_\mu = (\omega, 0, 0, - \omega)^T$, the above (anti-)commutators can be
expressed in explicit components of $a_\mu (k)$ as
$$
\begin{array}{lcl}
&& [ Q_{d}, a_{0}^\dagger (k) ] =
\bar c^\dagger (k), \;\;\;\qquad\;\;\;
[ Q_{d}, a_{0} (k) ] = \bar c (k), \nonumber\\
&& [ Q_{d}, a_{3}^\dagger (k) ] = -
\bar c^\dagger (k), \;\;\;\qquad\;\;\;
[ Q_{d}, a_{3} (k) ] = -
\bar c (k), \nonumber\\
&& [ Q_{d}, a_{1}^\dagger (k) ] = 0, \;\;\;\;\qquad\;\;\;\;
[ Q_{d}, a_{1} (k) ] = 0, \nonumber\\
&&[ Q_{d}, a_{2}^\dagger (k) ] = 0, \;\;\;\;\qquad\;\;\;\;
[ Q_{d}, a_{2} (k) ] = 0, \nonumber\\
&& \{ Q_{d}, c^\dagger (k) \} =  i
\bigl ( a_{0}^\dagger (k) +
a_{3}^\dagger (k) \bigr ), \qquad
\{ Q_{d}, \bar c^\dagger (k) \} =  0,
\nonumber\\
&& \{ Q_{d}, c (k) \} =  i
\bigl ( a_{0} (k) +
a_{3} (k) \bigr ), \qquad
\{ Q_{d}, \bar c (k) \} =  0.
\end{array} \eqno(5.7)
$$
The above (anti-)commutators between $Q_{d}$ and the creation and annihilation
operators can be re-expressed in the covariant form by exploiting the choice
of $k_\mu = (\omega, 0, 0, -\omega )^T$ and $k^\mu = (\omega, 0, 0, \omega)^T$ as
$$
\begin{array}{lcl}
&& [ Q_{d}, a_{\mu}^\dagger (k) ] =
{\displaystyle \frac{1}{\omega}} \;k_\mu \bar c^\dagger (k),
\qquad [ Q_{d}, a_{\mu} (k) ] =
{\displaystyle \frac{1}{\omega}}\; k_\mu \bar c (k), \nonumber\\
&& \{ Q_{d}, \bar c^\dagger (k) \} = 0, \;\;\;\;\;\qquad \;\;\;\;\;
\{ Q_{d}, \bar c (k) \} = 0, \nonumber\\
&& \{ Q_{d}, c^\dagger (k) \} =   {\displaystyle \frac{i}{\omega}}\;
k^\mu a_{\mu}^\dagger (k), \qquad
\{ Q_{d}, c (k) \} =
{\displaystyle \frac{i}{\omega}}\; k^\mu a_\mu (k).
\end{array} \eqno(5.8)
$$
The dual gauge transformation on the polarization vector
(i.e. $ e_\mu (k) \rightarrow (e_\mu)^{(dg)} (k) = e_\mu (k) +
i B k_\mu $  where B is a complex number which can be expressed
in view of the equations (4.10) and (4.11) as
$ B = (u e^1 + v e^2)/ (i)$)  will correspond to the
following expression in the language of the co-BRST charge $Q_{d}$
$$
\begin{array}{lcl}
| e + i\; B k, vac > = | e, vac > + \;Q_{d} \;(B\; \omega\;
c^\dagger (k)) \;| vac >, \qquad Q_{d}\; |vac> = 0.
\end{array} \eqno(5.9)
$$
Here we have used the anti-commutator $\{ Q_{d}, c^\dagger (k)\} =
(i/\omega) k^\mu a_\mu^\dagger (k)$ from (5.8) and invoked the
condition $Q_{d} |vac> = 0$ (cf equation (5.3)). It is worthwhile to mention
that the derivation of (5.9) is not dependent on the  specific choice of
the reference frame. A general proof of (5.9) can be given by exploiting
the general (anti-)commutators of (5.6). It is clear, from (5.6), that
the following anti-commutation relation is correct, namely;
$$
\begin{array}{lcl}
\{ Q_{d}, k_{0}\; c^\dagger (k) \} \;| vac >
= i (k_0 a^\dagger _{0} - k_i a^\dagger_i) \;|vac>
= i k^\mu a_\mu^\dagger (k) |vac> \equiv i\;| k, vac>,
\end{array} \eqno(5.10)
$$
where we have used the masslessness ($ k^2 = 0$) condition which implies
$k_{0}^2 = {\bf k}^2$. Thus, in the general case, the analogue of (5.9) is
$$
\begin{array}{lcl}
| e + i\; B k, vac > = | e, vac > + \;Q_{d} \;(B\; k_{0}\;
c^\dagger (k)) \;| vac >, \qquad Q_{d}\; |vac> = 0.
\end{array} \eqno(5.11)
$$
The above equations (5.9) and (5.11) imply that the dual-gauge transformed 
state in the quantum Hilbert space is
equal to the sum of the original state and a BRST co-exact state.
With the four nilpotent and conserved charges $Q_{(a)b}, Q_{(a)d}$ and
a  bosonic conserved charge $Q_{w}$ in the theory, the most symmetric
physical state ($| phy >$) can be defined as
$ Q_{(a)b} | phy > = 0, Q_{(a)d} | phy > = 0, Q_{w} | phy > = 0.$
Applying this physicality condition on the single photon state, we obtain the
following relationships by exploiting the commutators
$ [ Q_{b}, a_\mu^\dagger (k) ] = k_\mu c^\dagger (k),
[Q_{d}, a_\mu^\dagger (k)] = (k_\mu/\omega)\; \bar c^\dagger (k)$ of
equations (5.2) and (5.8), namely;
$$
\begin{array}{lcl}
Q_{b} \;| e + i\; A \;k,  vac> &=& Q_{b} \;| e, vac > \equiv (k \cdot e)\;
c^\dagger (k)\; | vac> = 0, \qquad
(Q_{b}^2 = 0),\nonumber\\
Q_{d} | e + i\; B \;k,  vac> &=& Q_{d} \;| e, vac > \equiv
{\displaystyle \frac{(k \cdot e)}{\omega}}
\; \bar c^\dagger (k)\; |vac> = 0, \qquad
(Q_{d}^2 = 0),
\end{array} \eqno(5.12)
$$
which imply the {\it transversality} (i.e. $ k \cdot e = 0$)
of the photon because of the fact
that $ c^\dagger (k) | vac > \neq 0, \bar c^\dagger (k) |vac> \neq 0$. It is
also obvious from the above discussion that for a single photon state,
{\it not} satisfying the above transversality condition, the single particle
(anti-)ghost state(s) ($\bar c^\dagger (k) |vac>, c^\dagger (k) |vac>$)
created by the operators $\bar c^\dagger (k)$ and $c^\dagger (k)$ would
turn out to be the BRST (co-)exact states. This explains the 
{\it no-(anti-)ghost}
theorem in the context of the BRST cohomology. Physically, it amounts to
the well-known fact (see, e.g., [32]) that the
contributions coming from the longitudinal and
scalar degrees of freedom of
the photons are cancelled by the presence of (anti-)ghost fields. 
This statement
is valid at any arbitrary order of perturbative calculations.
Ultimately, the physicality
criteria $Q_{b} | e, vac> = 0, Q_{d} |e, vac> = 0, Q_{w} |e, vac> = 0$
on a single photon state implies the transversality and masslessness
of the photon. \\

\noindent
{\bf 6 Conclusions}\\

\noindent
In the present investigation, we have been able to demonstrate a deep
connection between the transformations on the polarization vector $e_\mu (k)$
generated by (i) the translation subgroup $T(2)$ of the Wigner's little
group (ii) the $U(1)$ group of gauge symmetry, and (iii) the dual version
of the $U(1)$ gauge symmetry. It turns out that the (dual-)gauge symmetries 
owe their origin to the Wigner's little group. The connection between the
dual-gauge transformations and the Wigner's little group is a new
observation which has also been found for the 2-form free Abelian gauge theory
in 4D [21]. It is worthwhile to point out that the non-local
and non-covariant dual gauge symmetries look quite trivial when we choose
the reference frame where the momentum vector $k^\mu$, for a propagating
massless photon with energy $\omega$
along the z-direction of the 4D Minkowskian flat manifold, takes the form
$k^\mu = (\omega, 0, 0, \omega)^T$. 
In fact, the virtues of this choice of the reference frame are three fold.
First, the non-locality and non-covariance vanish from the co-BRST 
(or dual gauge)
symmetry transformations. Second, it enables us to tackle the subtleties 
and intricacies
associated with the discussions on Wigner's little group in a much 
nicer and better way.
Finally, all the (anti-)commutators  between the co-BRST charge $Q_{d}$ and
the creation and annihilation operators become very simple.
As a consequence of this simplicity, the physicality criteria
with the (co-)BRST charges (i.e. $ Q_{(d)b} |phys> = 0$)
on the physical harmonic state lead to the same physical inferences on
the 4D photon. That is to say, the 4D photon is found to be massless
($ k^2 = 0$) and transverse ($ k \cdot e = 0$) due to both the conserved
and nilpotent (co-)BRST charges $Q_{(d)b}$. However, a closer look at the
physicality conditions (5.12) sheds some light on an important difference
between the physical inferences drawn from the (co-)BRST charges. Whereas
the BRST charge leads to {\it only} the transversality
($ k \cdot e = 0$) condition, the co-BRST charge implies (i) the
transversality ($k \cdot e = 0)$ condition for a photon with finite energy
$\omega$, and (ii) the transversality condition might not be absolutely
essential condition for a highly energetic ($ \omega \rightarrow \infty$)
photon. In contrast, it would be worthwhile to mention
that, the physicality criteria with the (co-)BRST charges
for the one-form gauge theory 
in 2D [16,17]
(and 2-form gauge theory in 4D [21]), lead to mathematically 
{\it two} different kinds of relationships
between the momentum vector $k_\mu$ and the polarization vector $e_\mu$  
of the one-form gauge theory
(and the polarization anti-symmetric tensor $e_{\mu\nu}$ of the 2-form 
gauge theory).
As far as the discussion on the BRST cohomology is concerned, we have been
able to show that the (dual-)gauge (or (co-)BRST) transformed states are
the sum of the original states and the (co-)BRST exact states. Thus, the 
genuine
physical states of the theory remain invariant under the (dual-)gauge
(or (co-)BRST) transformations because the increment in the physical
states, due to the above transformations, turns out to  be cohomologically
trivial state which does not change the physical contents in any way.

It would be interesting endeavour to capture these 4D
(dual-)gauge (or (co-)BRST) symmetries in the framework of superfield
formulation where the geometrical origin for the nilpotent charges can
be found out. In fact, for the 2D free Abelian and self-interacting non-Abelian
gauge theories, such studies have already been performed [34-39] where the 
super
de Rham cohomological operators $(\tilde d, \tilde \delta, \tilde \Delta)$
have been exploited in the (dual-)horizontality conditions. Moreover, the
topological properties of these theories have also been encompassed in
the framework of the superfield formalism developed on the four $(2 + 2)$
dimensional manifold where the Lagrangian density and the symmetric
energy momentum tensor of the theory have been shown to correspond to
the translation of some composite superfields along the Grassmannian
directions. Thus, the superfield formulation of the (co-)BRST symmetries
with (non-)local transformations, is a key direction that should
be pursued in our further investigations. It is interesting to point 
out that, for the 4D
Abelian gauge theory, these (non-)local and (non-)covariant  
transformations corresponding to the (co-)BRST symmetries have been
recently captured in the superfield formulation where the geometrical
origin and interpretation for these symmetries have been provided
(see, e.g., [40] for details). We have not discussed, in the present
work, the non-local and non-covariant symmetry transformations
for the {\it interacting} theory  where there is a coupling between
the $U(1)$ gauge field and the Noether conserved current constructed
by the matter (Dirac) fields. These transformations for the gauge
as well as matter fields have
been discussed in detail (see, e.g., [26,31,40] and references therein).

It is an open problem to capture the symmetry transformations on the 
Dirac (matter) fields in the framework
of the Wigner's little group and/or the superfield formulation. Some
thoughts are being given to this problem at the moment. In this
connection, it is gratifying to state that the local, covariant, continuous
and nilpotent (anti-)BRST transformations for the matter fields have 
recently been derived in the framework of superfield formalism [41,42]
for the 2D and 4D interacting gauge theories. The local, covariant, continuous
and nilpotent (anti-)co-BRST symmetries for the Dirac fields in the 
case of 2D QED have also been derived (see, e.g., [41] for details).
However, there is no clue, so far, on such derivations (for the
matter fields) in the framework of the Wigner's little group. The derivation
of the gauge symmetry transformations for the non-Abelian gauge field
is an open problem in the context of the Wigner's little group. It is
obvious from our earlier works [20,21] that the Wigner's little group
does shed some new light on the structure of the higher-spin
(e.g. 2-form) gauge field theories in the framework of 
BRST formalism. Thus, we hope that such studies for still higher-spin
(e.g. 3-form, 4-form, etc.) gauge fields will be relevant to the (super)string
theories and the related areas of the
extended objects (e.g. D-branes, membranes, etc.) which are of interest
at the frontier level of research devoted to the discussions
of physics at Planck energy. The above 
directions are some of the key issues that are under investigation at the 
moment and our results will be reported elsewhere [43].\\

\noindent
{\bf Acknowledgements}\\

\noindent
My present work is dedicated to the memory of my eldest brother
Pt. J. P. Malik who is no more in this world but
his interest in science and his passion for the vocal
classical (Hindustani) music will remain a perpetual source of inspiration for
all my scientific activities and other endeavours in life. A crucial and very 
important issue of this work was settled and thoroughly fixed at the AS-ICTP, 
Trieste, Italy. The warm hospitality
extended to me by the HEP group (at AS-ICTP) is gratefully acknowledged.

\end{document}